 \documentclass[12pt,preprint]{aastex}

 \newcommand{\kms}{~km s$^{-1}$~}
\newcommand{\myemail}{zhekovs@colorado.edu}

\shorttitle{Shock Kinematics in SNR~1987A} \shortauthors{Zhekov et al.}

\begin{document}

%% LaTeX will automatically break titles if they run longer than % one line.
%However, you may use \\ to force a line break if % you desire.

\title{Chandra Observations of Shock Kinematics in Supernova Remnant 1987A}

%% Use \author, \affil, and the \and command to format % author and
%affiliation information.  % Note that \email has replaced the old
%\authoremail command % from AASTeX v4.0. You can use \email to mark an email
%address % anywhere in the paper, not just in the front matter.  % As in the
%title, use \\ to force line breaks.

%\author{Svetozar A. Zhekov\altaffilmark{1,4}, 
%Richard McCray\altaffilmark{1},
%Kazimierz J. Borkowski\altaffilmark{2},
%David N. Burrows \altaffilmark{3}, and Sangwook Park\altaffilmark{3}}

\author{S.A. Zhekov\altaffilmark{1,4},
R. McCray\altaffilmark{1},
K.J. Borkowski\altaffilmark{2},
D.N. Burrows \altaffilmark{3}, and S. Park\altaffilmark{3}}

%% Notice that each of these authors has alternate affiliations, which % are
%identified by the \altaffilmark after each name.  Specify alternate %
%affiliation information with \altaffiltext, with one command per each %
%affiliation.

\altaffiltext{1}{JILA, University of Colorado, Boulder, CO 80309-0440; 
\myemail, dick@jila.colorado.edu}
\altaffiltext{2}{Department of Physics, NCSU, Raleigh, NC 27695-8202;
kborkow@unity.ncsu.edu}
\altaffiltext{3}{Department of Astronomy and Astrophysics, Pennsylvania 
State University, 525 Davey Laboratory, University Park, PA 16802;
burrows@astro.psu.edu; park@astro.psu.edu}
\altaffiltext{4}{On leave from Space Research Institute, Sofia, Bulgaria}

%% style, abstract will output a Received/Accepted line after the % title and
%affiliation information. No date will appear since the author % does not
%have this information. The dates will be filled in by the % editorial office
%after submission.

\begin{abstract}   We report the first results from deep X-ray observations 
of the SNR~1987A with the Chandra LETG. Temperatures inferred
from line ratios range from $\sim 0.1 - 2$ keV and increase with
ionization potential.  Expansion velocities inferred from X-ray line profiles
range from $\sim 300 - 1700$ \kms, much less than the velocities inferred
from the radial expansion of the radio and X-ray images.  We can account for
these observations with a scenario in which the X-rays are emitted by shocks
produced where the supernova blast wave strikes dense protrusions of the
inner circumstellar ring, which are also responsible for the optical hot
spots.  \end{abstract}

%% Keywords should appear after the \end{abstract} command. The uncommented %
%example has been keyed in ApJ style. See the instructions to authors % for
%the journal to which you are submitting your paper to determine % what
%keyword punctuation is appropriate.

%% Authors who wish to have the most important objects in their paper %
%linked in the electronic edition to a data center may do so in the % subject
%header.  Objects should be in the appropriate "individual" % headers (e.g.
%quasars: individual, stars: individual, etc.) with the % additional
%provision that the total number of headers, including each % individual
%object, not exceed six.  The \objectname{} macro, and its % alias \object{},
%is used to mark each object.  The macro takes the object % name as its
%primary argument.  This name will appear in the paper % and serve as the
%link's anchor in the electronic edition if the name % is recognized by the
%data centers.  The macro also takes an optional % argument in parentheses in
%cases where the data center identification % differs from what is to be
%printed in the paper.

%\keywords{supernova remnants: --- supernovae: general --- supernovae:
%individual (\objectname{SNR 1987A}) --- X-rays: ISM }

%% From the front matter, we move on to the body of the paper.  % In the
%first two sections, notice the use of the natbib \citep % and \citet
%commands to identify citations.  The citations are % tied to the reference
%list via symbolic KEYs. The KEY corresponds % to the KEY in the \bibitem in
%the reference list below. We have % chosen the first three characters of the
%first author's name plus % the last two numeral of the year of publication
%as our KEY for % each reference.

\section{Introduction} 
\label{sec:intro} 
With the rapidly developing impact
of the debris of Supernova 1987A with its inner circumstellar ring, we have
an unprecedented opportunity to witness the birth of a supernova remnant,
SNR1987A (McCray 2005).  This event, the first hint of which was the 1995
appearance of Spot~1, a rapidly brightening optical `hotspot' on the ring,
has now evolved to the stage that the ring is encircled by hotspots (Sugerman
et al. 2002).  
Additionally, an annular source of X-ray emission
correlated with the locations of the hotspots has brightened at an
accelerating rate (Park et al. 2004).
Evidently, the optical hotspots appear where the supernova blast wave
encounters fingers of relatively dense gas protruding inwards from the ring.
In such a situation, a complex hydrodynamic interaction ensues (cf.
Borkowski, Blondin \& McCray 1997a,b). The optical emission from the spots
comes from relatively slow ($V_S \le 200$~km s$^{-1}$) shocks which have had
time to undergo radiative cooling 
(Pun et al. 2002).  The X-ray
emission must come from gas heated by faster ($V_S \ge 1000$~km s$^{-1}$)
shocks, either transmitted shocks entering the protrusions or shocks
reflected from the protrusions.

Michael et al. (2002) reported the first observation of a dispersed X-ray
spectrum of SNR~1987A, taken in October 1999 with 
HETG on the {\it Chandra} observatory.
The X-ray emission was dominated by shock-heated gas having
electron temperature $kT_e \approx 2.6$ keV.
Due to the poor photon statistics, only a composite line profile was 
constructed by stacking the profiles of the
individual observed lines.
From the measured FWHM ($\approx 5000$~km
s$^{-1}$), they inferred that the X-ray emitting gas was moving with radial
velocity $\approx 3500$~km s$^{-1}$, roughly the same as that inferred from
the observed proper motion of the non-thermal radio source (Manchester et al.
2002) and the X-ray source (Park et al. 2004).

From October 1999 to September 2004, the X-ray source SNR1987A has
brightened by a factor $\sim 10$ (Park et al. 2005). As a result, it has
become possible with {\it Chandra} to obtain dispersed X-ray spectra with
very high counting statistics.
In this {\it Letter}, we report the first results from the 
analysis of such observations.

\section{Observations and Data Reduction }
\label{sec:obs}

SNR~1987A was observed with {\it Chandra} in the configuration LETG-ACIS-S in
five consecutive runs during Aug 26 -- Sep 5, 2004, providing a total
effective exposure of 289 ksec. 
We extracted\footnote[1]{For CIAO 3.1 and ATOMDB 
see http://cxc.harvard.edu/ciao/ and http://cxc.harvard.edu/atomdb/ }
 positive and negative
%CIAO 3.1\footnote[1]{see http://cxc.harvard.edu/ciao/} was used
%to extract the 
first-order LETG spectra for each of the five observations. 
Then, we merged
the resultant spectra into one spectrum each for the positive and negative
LETG arms with respective total counts of 9,241 and 6,057 in the energy range
0.4 - 7 keV. 
The difference in photon statistics is a result of the different
sensitivities of the respective CCD detectors.  We also extracted the
pulse-height spectrum from the zeroth-order image with a total number of
16,557 counts in the 0.4 - 7 keV range.
Figure~\ref{fig:spec} demonstrates the enormous scientific advantage
of  the dispersed spectrum over the pulse-height spectrum. 

We fitted the strong emission line triplets of various He-like ions
%(and Fe XVII as well)
by a sum of three Gaussians and constant local continuum.
The ratio of line centers of the triplet components were held fixed 
according to the values given by the {\it Chandra} atomic
database$^1$
%database\footnote[1]{see http://cxc.harvard.edu/atomdb/},
and all components shared the same full width at half maximum (FWHM).
Thus, 
free parameters were the intensities of their components, the line center 
wavelength of one of the components, FWHM, and the local continuum. 
Likewise, we fitted the strong emission doublets of the H-like species
but the component intensities were fixed through their atomic data values.
%
%We fitted the strong emission doublets of the H-like species of various
%elements by a sum of two Gaussians and constant local continuum.  Free
%parameters were the intensity of the stronger component of the doublet, its
%line center wavelength, its full width at half maximum (FWHM), and the
%continuum level.  The second component was assumed to have the same FWHM. The
%displacement of its line center wavelength and the ratio of its intensity to
%the stronger component were held fixed according to the expected values given
%by the {\it Chandra} atomic database\footnote[1]{see
%http://cxc.harvard.edu/atomdb/}.
%Likewise, we fitted the strong emission line
%triplets of various He-like ions but the intensities of all three
%components were free parameters of the fit.
We found that the centroid shifts for the strong
emission lines were consistent with the red-shift of the Large Magellanic
Cloud, and that the line fluxes derived from the positive and negative
first-order spectra agreed within expected statistical uncertainties.
Therefore, we assumed that all the line centers had the same Doppler shift,
$V_D = 286$\kms, and fitted the positive and negative first-order spectra
simultaneously.

\section{Line Fluxes and Ratios}

Table~\ref{tab:flux} lists the fluxes of all the emission lines and
multiplets that could be measured with acceptable photon statistics.  Given
the excellent spectral resolution and photon statistics, we can for
the first time derive reliable intensities and ratios of various X-ray
emission lines from SNR~1987A.  As we shall discuss in \S~\ref{sec:disc}, we
expect the actual conditions in the plasma to be sufficiently complex that we
can only use the line strengths and ratios to infer typical conditions of the
plasma responsible for 
%most of the emission by a given ion. 
the emission by a given ion. 
The ratio of 
%K$_\alpha$/Ly$_\alpha$ 
%K$_\alpha$(He-like)/Ly$_\alpha$(H-like) 
 He-like(K$_\alpha$)/H-like(Ly$_\alpha$)
from a given element is sensitive to both
temperature and ionization state of the gas. 
%Likewise, the `G-ratio' of the
%Likewise is the `G-ratio' of the
The same is true for the `G-ratio' of the
He-like triplet lines ($G = \frac{f+i}{r}$~where `f', `i' and `r' stand for
forbidden, intercombination and recombination lines, respectively) 
%is sensitive to both temperature and ionization state, 
since these lines are
produced not only by electron impact excitation of the He-like ions but also
by K-shell ionization of the Li-like ions (Mewe 1999; Liedahl 1999).  

Since we expect that shocks are responsible for heating and ionizing the
X-ray emitting gas, we can compare our measured line ratios to those
resulting from the XSPEC code, which provides models of the time-dependent
ionization and X-ray emission from plane-parallel shocks (Borkowski, Lyerly,
\& Reynolds 2001). In such models, the line ratios are functions of the
post-shock temperature and the ionization age, $n_et$, defined as the time
since the gas first entered the shock times the postshock electron density.
The X-ray emitting plasma in SNR~1987A is in NEI (nonequilibrium ionization),
and so inner-shell ionization and excitation contribute importantly to
emission line ratios.  
%emission line ratios (Mewe 1999; Liedahl 1999).  
Indeed, if these processes are not included, 
the theoretical G-ratios cannot match the
observed ones. 

In Figure~\ref{fig:ratios}, 
the curves define those
regions in the parameter space of electron temperature and ionization age,
$n_et$, for which the observed line ratios agree with the ratios that would
result from a plane parallel shock. 
Except for a very narrow range of $n_et$, the inferred temperatures
are likely to be in the range 0.1--2~keV.
We see immediately that no single
combination of electron temperature and ionization age is consistent with all
the observed ratios. Instead, for any given ionization age, the inferred
electron temperature increases 
%systematically 
tentatively
with ionization potential.
Evidently, the X-ray emission
comes from a distribution of shocks having a range of ionization ages and
post-shock temperatures.

\section{Line Profiles} 
\label{sec:profiles} 

\begin{figure} 
\begin{center} 
\includegraphics[width=2.25cm, height=8.5cm,angle=-90]{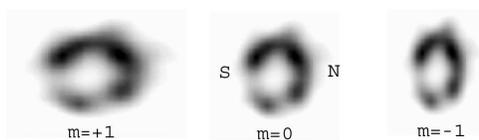}
\end{center} 
\caption{A schematic view of the LETG observational situation for SNR~1987A
using the deconvolved zeroth-order total image. 
The X-ray source is assumed to be an
inclined ring, the northern (N) side of which is blue-shifted and the
southern (S) side of which is red-shifted. Shown are the zeroth-order image
($m = 0$), as well as the positive ($m = +1$) and negative ($m = -1$) images
in an arbitrary  spectral line.  The dispersion direction is horizontal.  }
\label{fig:cartoon} 
\end{figure}

The excellent spatial and spectral resolution of {\it Chandra} allows us for
the first time to observe the kinematics of the X-ray emitting gas through
the line profiles.  The method is illustrated by Figure
\ref{fig:cartoon}.\footnote[2]{For the general properties of LETG
see \S 9 of the Chandra Proposer's Observatory Guide, Version 7.0, 
p.195; Complexity of the spatial-spectral effects is discussed in
\S~8.5.3, pp.187-189; \S~9.3.3, pp.209-215.} 
We assume that the X-ray source lies on a circular ring having
angular radius $\theta_R$ and that this ring is expanding with constant
radial velocity, $V_R$.  
We also assume that the X-ray source has, 
like the optical
inner circumstellar ring, inclination angle $i = 44-45^\circ$, with the near
side to the north and minor axis at P.A. $\approx 354^\circ$
~(Sugerman et al. 2002).
The roll angle was chosen so that the negative ($m = -1$)
arm of the dispersion axis was aligned at P.A. $\approx 345^\circ$,
thus, the north side of the ring will be blue-shifted and the south
side will be red-shifted.
The dispersed images of the ring will be distorted by these 
Doppler shifts.$^2$
In the $m = -1$ image, the N
side of the ring will be displaced to the left, and the S side will be
displaced toward the right. Thus, the minor axis of the $m = -1$ image will
be compressed.  Likewise, the minor axis of the $m = +1$ image will be
stretched.

%, very close
%to the minor axis of the inner circumstellar ring (P.A. $\approx 354^\circ$).
%thus, the north side of the ring (on the right in Fig. \ref{fig:cartoon})
%will be blue-shifted (displaced toward the zero-order image) and the south
%side will be red-shifted (away from zero-order).  The dispersed images of the
%ring will be distorted by these Doppler shifts.  In the $m = -1$ image, the N
%side of the ring will be displaced to the left, and the S side will be
%displaced toward the right. Thus, the minor axis of the $m = -1$ image will
%be compressed.  Likewise, the minor axis of the $m = +1$ image will be
%stretched.

\begin{figure} 
\begin{center} 
\includegraphics[width=9cm, height=6cm]{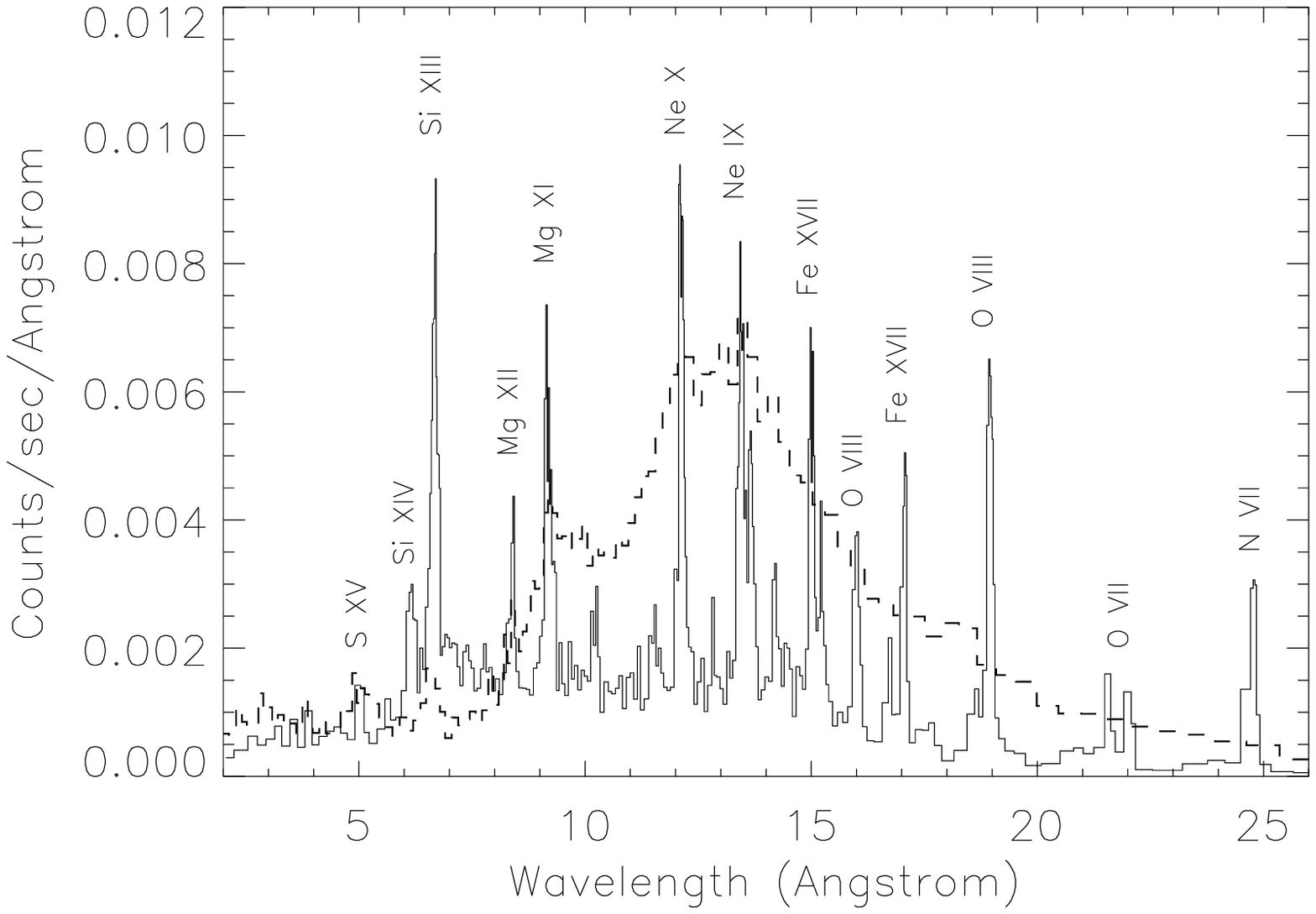} 
\includegraphics[width=7cm, height=6cm]{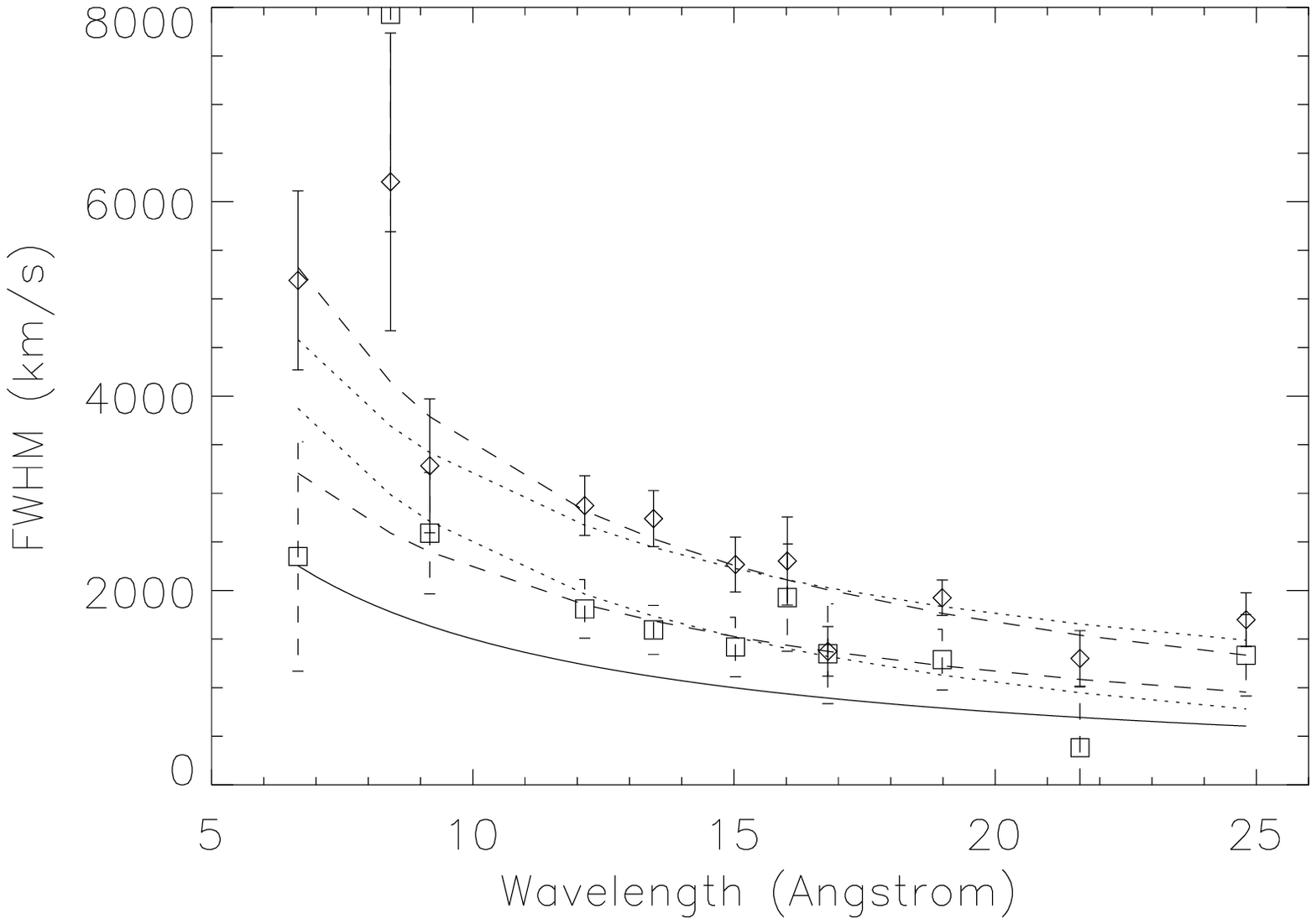}
\end{center} 
\caption{{\bf Left panel:}
The background-subtracted LETG
first ($+1$, solid) and zeroth-order (dashed) spectra.
{\bf Right panel:}
Measured line widths (FWHM) for the positive ({\it diamonds}) and negative
({\it squares}) LETG arms, respectively.  The solid curve represents the
resolving power of the LETG 
($E/\Delta~E \approx 20\lambda$).  The dashed and dotted
curves correspond to the line broadening parameter in the cases with and
without shock stratification (see \S~\ref{sec:profiles}). }
\label{fig:spec}
\end{figure}

This behavior is exactly what we see in the measured line profiles. We fitted
Gaussian profiles 
(as explained in \S~\ref{sec:obs}) to
each emission line or multiplet independently in the $m = +1$ and $m = -1$
first-order spectra.  
In every case, the width of a given emission line in the
$m = +1$ arm is greater than the corresponding width in the $m = -1$ arm
(Fig.~\ref{fig:spec}). 
%We note that the actual line profile may differ from a Gaussian due to the
%complex spatial-spectral effects. However, our choice is justified
%by the quality of the line-profile fits (reduced $\chi^2 \approx 1$) and 
%a more elaborated approach (based on the MARX software) 
%will be presented elsewhere.
%
The actual line profile will likely differ from a Gaussian due to the
complex spatial-spectral effects. However, the Gaussian approximation 
is sufficient for the analysis presented here
(resulting in a reduced $\chi^2 \approx 1$ for each of the line-profile
fits).

There are two major sources of line broadening in the dispersed images.
One is
the spatial extent of the image itself, which can be expressed as an
equivalent line broadening, independent of wavelength.  
The second is the broadening due to
the bulk motion of the shocked gas. We express the net line width (FWHM)
broadening as: 
\begin{equation} 
\Delta \lambda_{tot} = 2 \Delta \lambda_0 \pm 2 z_0
(\lambda/\lambda_0)^\alpha \lambda\,,~ 
\label{eqn:stratified} 
\end{equation}
where the plus (minus) sign refers to the $m = + 1$ ($m = - 1$) spectrum,
respectively. 
The two sources of broadening are represented respectively
 by the first and second term
on the right hand side of equation~(\ref{eqn:stratified}).
The power-law function of wavelength
allows for the possibility that the mean bulk velocity of shocked gas
emitting a given line may depend on the excitation or ionization stage of the
emitting ion.  The parameter $z_0$ determines the line broadening at some
fiducial wavelength, $\lambda_0$, and the power-law index $\alpha$ is to be
determined.

We then fit equation (\ref{eqn:stratified}) to the data in Figure
\ref{fig:spec}. We consider two models.  In the first (constant velocity)
model, we assume that all emitting ions have the same mean bulk velocity
($\alpha = 0$).  In that case, we find a best-fit value
$z_0 = 0.0006$. In the second (stratified) model, we allow the values of
$\alpha$ and $z_0$ 
to vary.  In that case, we find a best-fit
value $\alpha = -1.3$.  Note that such a negative value of $\alpha$ 
implies that
emission lines with shorter wavelength are produced by gas having greater
radial velocity, as expected.  
%It is evident from Figure \ref{fig:spec} that 
The stratified model provides a slightly better fit to
the data than the constant velocity model 
%($\chi^2/dof = 22/19$~vs. $25/20$~, see also Fig.\ref{fig:spec}), 
($\chi^2/dof = 22/19$~vs. $25/20$), 
but the data are not good
enough to distinguish between the models with confidence.
We may translate the fitting parameters into an equivalent
mean radial bulk velocity, $V_r$, of an expanding circular ring by adopting
an average value for the azimuthal $\phi$ ($\overline{\sin\phi} = 2/\pi~,
0\le\phi\le\pi/2$) and a value ($i = 45^\circ$) for the inclination of the
inner ring.  This yields a value $V_r = 397$~km s$^{-1}$.  
In the case of a strong radial shock with adiabatic index $\gamma = 5/3$, the
corresponding shock velocity must be $V_S = 4V_r/3 = 530$\kms.  The
corresponding shock velocity in the stratified case is given by
$V_{S}(\lambda) = 920~(\lambda/10 $\AA)$^{-1.3}$\kms.
Finally, we note that in the both models the derived source half-size of
$\Delta\lambda_0 = 0.047$~\AA~ or $\theta_R =0\farcs84$ is consistent with
the SNR~1987A size from an X-ray image-deconvolution technique (Burrows et
al. 2000). Thus, we are confident that the simplified treatment presented
here gives reliable results.

\section{Discussion and Conclusions} 
\label{sec:disc} 
\begin{figure} 
\begin{center} 
 \includegraphics[width=7.5cm, height=6.25cm]{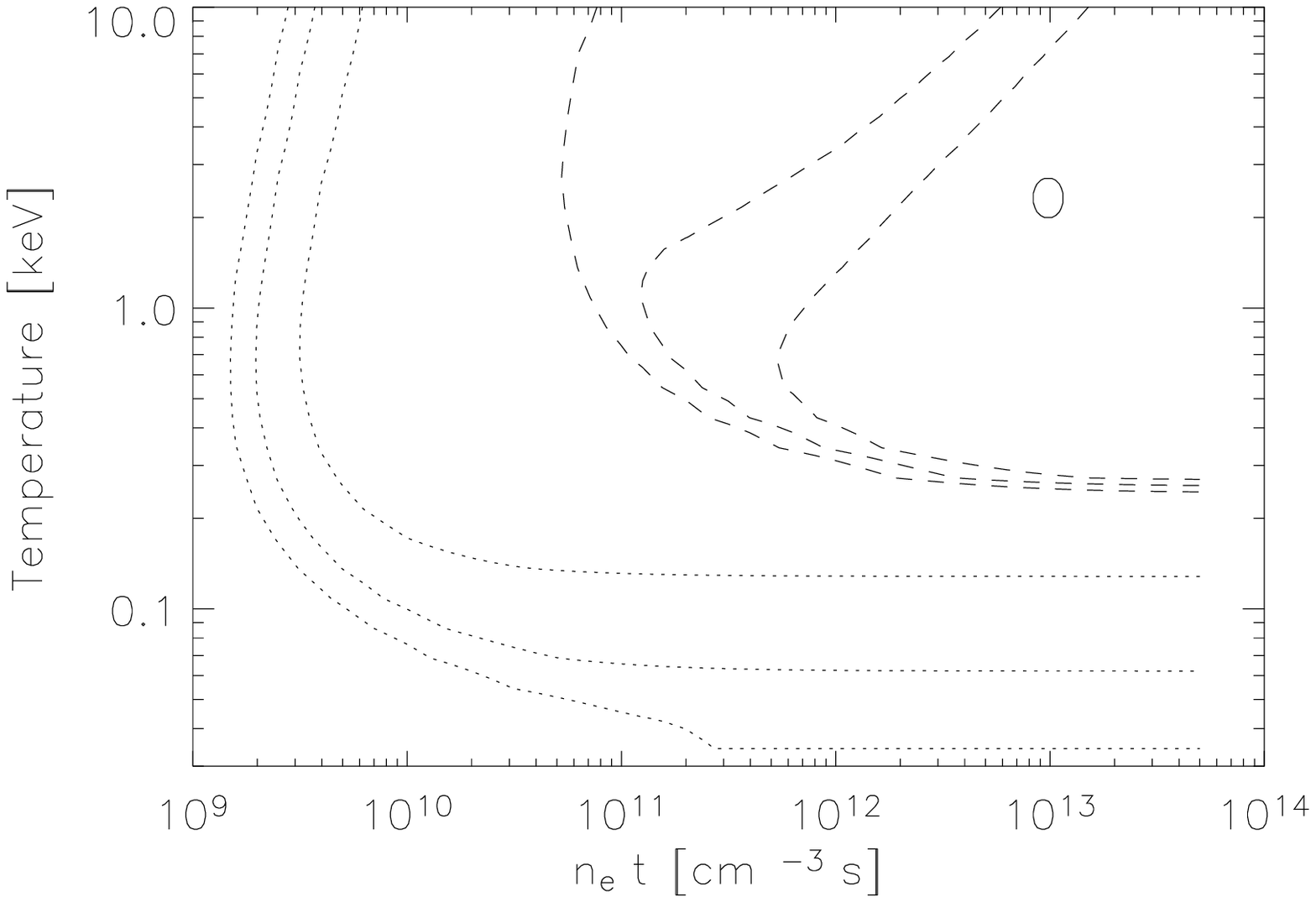} 
 \includegraphics[width=7.5cm, height=6.25cm]{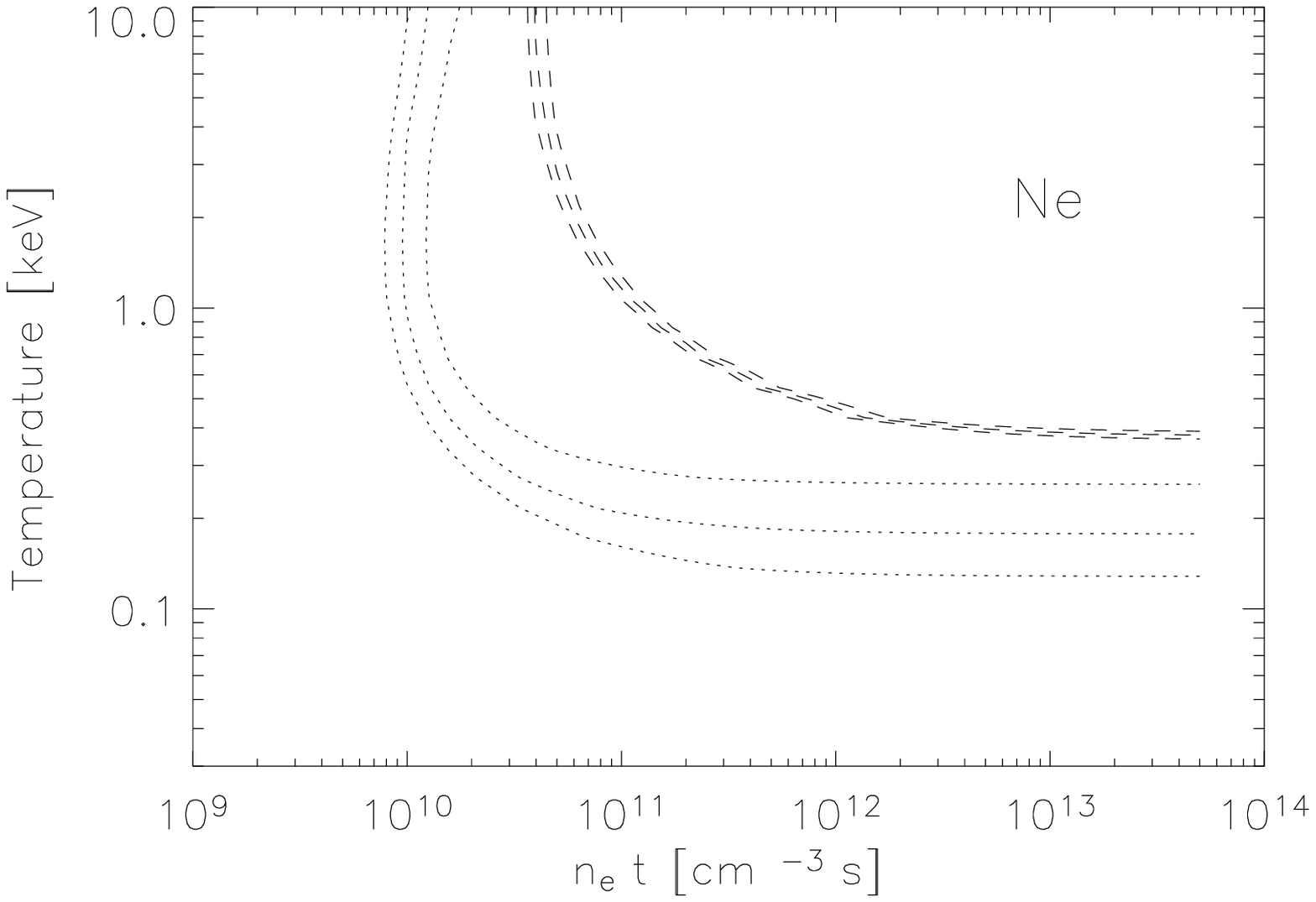}
 \includegraphics[width=7.5cm, height=6.25cm]{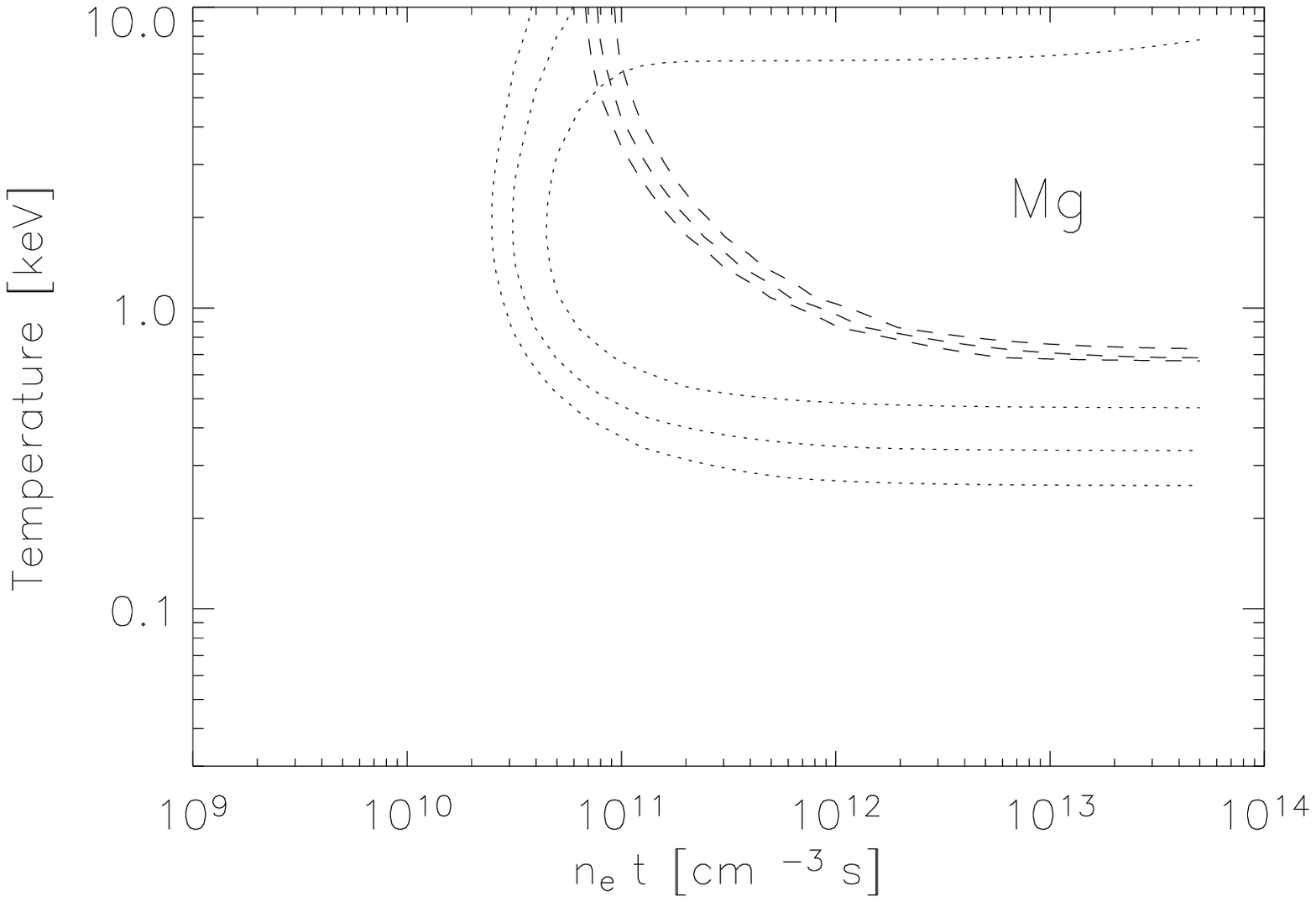}
 \includegraphics[width=7.5cm, height=6.25cm]{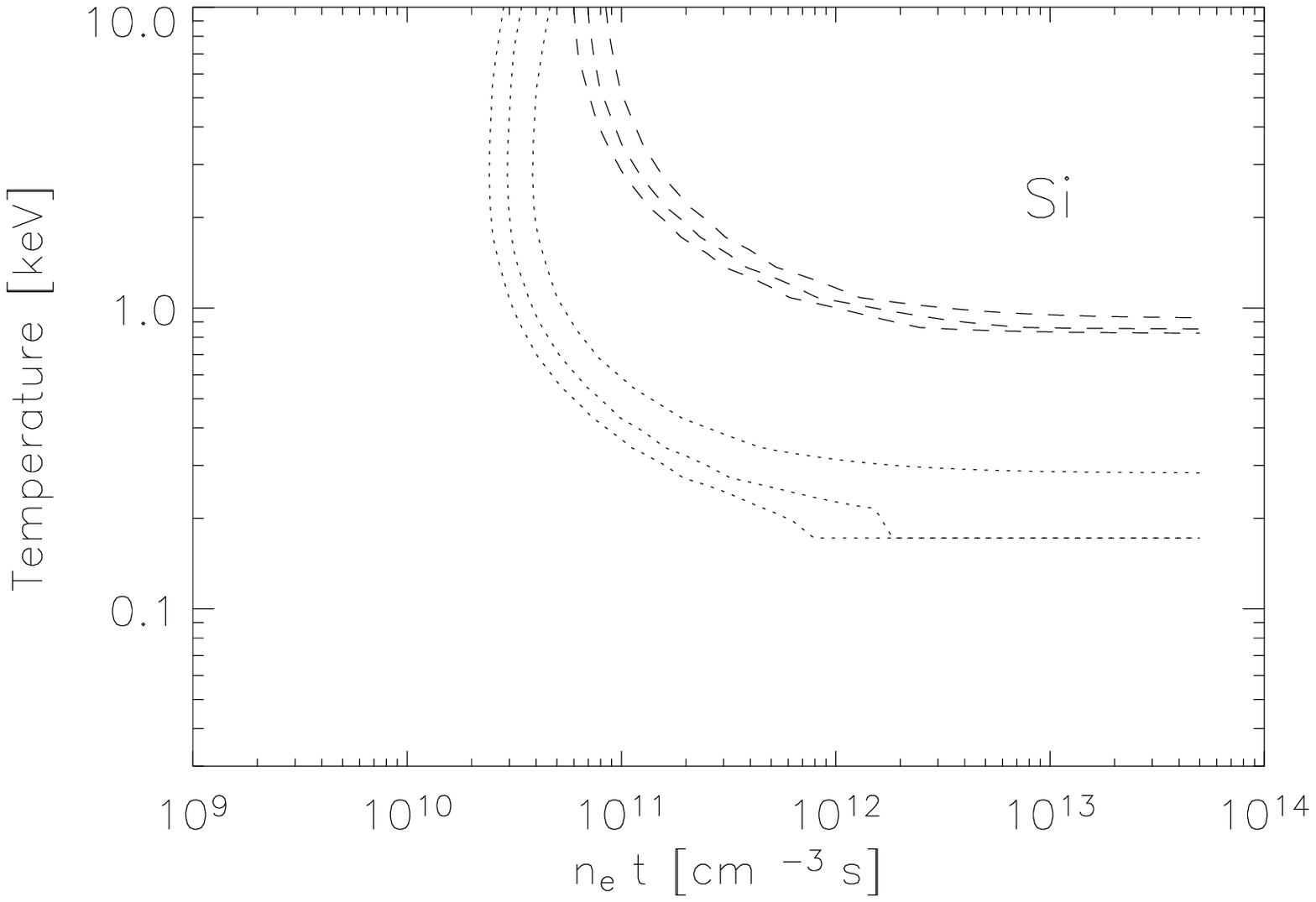} 
\end{center}
\caption{Isolines of the G-ratios (dots) and the
K$_{\alpha}$/L$_{\alpha}$ ratios (dashes) for O, Ne, Mg and Si. Note that the
G-ratios include the line contribution from inner-shell processes (see text).
The three curves for each ratio correspond to the measured value and $\pm
1\sigma$.  
The K$_{\alpha}$/L$_{\alpha}$ ratios have been corrected
for differential X-ray absorption assuming N$_H = 1.6\times10^{21}$~cm$^{-2}$. 
The observed G-ratios are: $1.44\pm0.47$~(O); $0.87\pm0.11$~(Ne);
$0.79\pm0.09$~(Mg); $1.17\pm0.23$~(Si).
} 
\label{fig:ratios} 
\end{figure}

The most surprising
result of this observation is the relatively low velocity of the X-ray
emitting gas as determined from the line profiles. When we proposed to do
this observation, we expected to see kinematic velocities in the range $\sim
2500 - 3000$\kms, as we saw in the composite line profile measured with the
{\it Chandra} HETG in October 1999 (Michael et al. 2002) and in the radial
expansion rate of the X-ray image (Park et al 2004).
At the very least, we would expect to see velocities comparable to the
velocity of gas behind a shock moving sufficiently fast to heat the electrons
to temperatures inferred from the X-ray line ratios.  For a shock of high
Mach number entering stationary gas with velocity $V_S$, the maximum electron
temperature is:
$kT_e = \frac{3}{16}\mu V_S^2 = 
%1.4 (V_S/1000~km s^{-1})^2\hspace{0.5cm}$~keV.
1.4 (V_S/1000~\mathrm{km~s^{-1}})^2\hspace{0.5cm}$~keV.
The value $V_S = 530$~km s$^{-1}$ inferred for the constant velocity model
then implies a post-shock 
temperature $kT_e \le 0.39$ keV.  This
value is inconsistent with the temperatures required to account for the
$K_\alpha/L_\alpha$  line ratios observed for Mg and Si
(Fig.~\ref{fig:ratios}).  
On the other hand, the electron
temperatures inferred from the line profiles in the stratified model may be
consistent with the observed line ratios ($kT_e = 0.15 - 4.0$~keV for $V_S =
340 - 1,700$\kms).  

How do we reconcile the relatively low gas velocities inferred from the line
profiles with the much greater velocities inferred from the radial expansion
rate of the X-ray image?
We have in mind a picture in which a blast wave of velocity $V_B \sim
3000$\kms propagates through the low atomic density ($n_0 \sim
100$~cm$^{-3}$) gas inside the circumstellar ring.  A zone of enhanced X-ray
emission appears when the blast wave strikes a finger of relatively dense
($n_1 \sim 3 \times 10^3 - 3 \times 10^4$~cm$^{-3}$) gas protruding inward
from the ring.  A transmitted shock propagates into the protrusion at
velocity reduced by a factor $\sim (n_1/n_0)^{1/2}$, while a reflected shock
propagates backwards.  
The latter increases the density and
temperature of the circumstellar matter to values substantially greater than
those caused by the original blast wave,
and it also reduces the radial velocity of the 
doubly-shocked circumstellar gas.  
The
enhanced X-ray emission comes from the gas behind the transmitted shock and
from the gas behind the reflected shock, with the proportional contributions
depending on the details of the hydrodynamics.
But, the lower velocity of the transmitted shock and the
high density behind it ensure a short cooling timescale
and eventually increased optical emission. 
At each
point of impact with a protrusion, a new zone of enhanced X-ray emission
appears.  Thus, the mean radius of the enhanced X-ray emission appears to
move outward at a substantial fraction of the blast wave velocity, while the
gas responsible for most of the X-ray emission may be moving much more
slowly.  This picture assumes that only a small fraction of the blast wave
area is covered by protrusions at the present time.  It also implies that the
radial expansion of the X-ray image should slow down rapidly as the blast
wave overtakes the entire equatorial ring.

The scenario that we have described here accounts in a natural way for
several properties observed by Chandra images and spectra of
SNR~1987A: (1) the correlation of the rapid X-ray brightening with the
appearance of the optical hotspots; (2) the correlation of the X-ray image
with the optical hotspots; (3) the relatively rapid expansion of the X-ray
image compared to the relatively slow bulk velocity of the X-ray-emitting
gas; and (4) the correlation of the inferred electron temperature and
expansion velocity of the shocked gas with excitation potential of X-ray
emission lines.  

There is much more to be learned from the Chandra spectroscopic data beyond
the brief summary that we have presented here.  By fitting global models for
the entire X-ray spectrum, we can infer element abundances and the
distribution of emission measure with temperature.  By simulating the actual
2-dimensional images in the dispersed spectra ({\it e.g.,} with the MARX
software), we can refine our models of the kinematics and spatial
distribution of the shocked gas.  By constructing simulations of the
hydrodynamics of the impact of the blast wave with the protrusions, we hope
to provide a more refined interpretation of the observations than the
analysis presented here.

%% The \notetoeditor{TEXT} command allows the author to communicate %
%information to the copy editor.  This information will appear as a %
%footnote on the printed copy for the manuscript style file.  Nothing will %
%appear on the printed copy if the preprint or % preprint2 style files are
%used.

%% The eqnarray environment produces multi-line display math. The end of %
%each line is marked with a \\. Lines will be numbered unless the \\ % is
%preceded by a \nonumber command.  % Alignment points are marked by
%ampersands (&). There should be two % ampersands (&) per line.

%% If you wish to include an acknowledgments section in your paper, %
%separate it off from the body of the text using the \acknowledgments %
%command.

%% Included in this acknowledgments section are examples of the % AASTeX
%hypertext markup commands. Use \url without the optional [HREF] % argument
%when you want to print the url directly in the text. Otherwise, % use either
%\url or \anchor, with the HREF as the first argument and the % text to be
%printed in the second.

\acknowledgments This work was supported by NASA through Chandra Awards
G04-5072A (to CU, Boulder, CO) and GO4-5072B (to NCSU, Raleigh, NC).
The authors appreciate the careful reading and comments by an
anonymous referee.

\clearpage

%% Here we use \plottwo to present two versions of the same figure, % one in
%black and white for print the other in RGB color % for online presentation.
%Note that the caption indicates % that a color version of the figure will be
%available online.  %

\clearpage

\clearpage

%% This figure uses \includegraphics to scale and rotate the still frame %
%for an mpeg animation.

%\begin{figure} \includegraphics[angle=90,scale=.50]{f3.eps}
%\caption{Animation still frame taken from \citet{kim03}.  This figure is
%also available as an mpeg animation in the electronic edition of the {\it
%Astrophysical Journal}.} \end{figure}

%% If you are not including electonic art with your submission, you may %
%mark up your captions using the \figcaption command. See the % User Guide
%for details.  % % No more than seven \figcaption commands are allowed per
%page, % so if you have more than seven captions, insert a \clearpage % after
%every seventh one.

%% Tables should be submitted one per page, so put a \clearpage before % each
%one.

%% Two options are available to the author for producing tables:  the %
%deluxetable environment provided by the AASTeX package or the LaTeX % table
%environment.  Use of deluxetable is preferred.  %

%% Three table samples follow, two marked up in the deluxetable environment,
%% one marked up as a LaTeX table.

%% In this first example, note that the \tabletypesize{} % command has been
%used to reduce the font size of the table.  % We also use the \rotate
%command to rotate the table to % landscape orientation since it is very wide
%even at the % reduced font size.  % % Note also that the \label command
%needs to be placed % inside the \tablecaption.

%% This table also includes a table comment indicating that the full %
%version will be available in machine-readable format in the electronic %
%edition.  %

\clearpage

\begin{table}
\begin{center}
\caption{SNR 1987A: Line Fluxes\label{tab:flux}}
\begin{tabular}{lrrrr}
\tableline\tableline
\multicolumn{1}{c}{Line} & \multicolumn{1}{c}{IP$^{a}$} &
\multicolumn{1}{c}{$\lambda_{lab}$$^{b}$} &
\multicolumn{1}{c}{Flux $^{c}$} & \multicolumn{1}{c}{Counts $^{d}$} \\
\tableline
S XV K$_{\alpha} $    & 3.22 &  5.0387  &  5.14 $\pm$ 2.15 & 87.5 $\pm$
36.6\\
Si XIV L$_{\alpha} $ & 2.67 &  6.1804  & 2.22 $\pm$ 0.46 & 102.7 $\pm$
21.3\\
Si XIII K$_{\alpha} $ & 2.44 &  6.6479  & 11.86 $\pm$ 1.19 & 522.7 $\pm$
52.4\\
Mg XII L$_{\alpha} $ & 1.96 &  8.4192  & 7.33 $\pm$ 0.96 & 250.6 $\pm$
32.8\\
Mg XI K$_{\alpha}$   & 1.76 & 9.1687  & 18.09 $\pm$ 1.11 & 475.7 $\pm$
29.2 \\
Ne X  L$_{\alpha} $  & 1.36 & 12.1321  & 36.40 $\pm$ 2.40 & 552.3 $\pm$
36.4\\
Ne IX K$_{\alpha}$   & 1.20 & 13.4473  & 60.68 $\pm$ 3.82  & 838.2 $\pm$
52.8\\
Fe XVII$^{e}$ & 1.27 & 15.0140  & 38.10 $\pm$ 3.20 & 450.4 $\pm$ 37.7\\
O VIII L$_{\beta} $ & 0.87 & 16.0055  & 19.90 $\pm$ 2.40 & 204.7 $\pm$
24.7\\
Fe XVII$^{e}$ & 1.27 & 16.7800  & 24.14 $\pm$ 3.18 & 336.4 $\pm$ 37.8\\
O VIII L$_{\alpha} $ & 0.87 & 18.9671  & 50.80 $\pm$ 4.00 & 377.3 $\pm$
29.7\\
O VII K$_{\alpha}$  & 0.74 & 21.6015  & 35.20 $\pm$ 5.77  & 162.8 $\pm$
26.7\\
N VII L$_{\alpha} $  & 0.67 & 24.7792  & 44.90 $\pm$ 5.20 & 203.7 $\pm$
23.6\\
\tableline
\end{tabular}
%% Any table notes must follow the \end{tabular} command.
\tablenotetext{a}{Ionization potential in keV.  }
\tablenotetext{b}{The laboratory wavelength of the main component
 in Angstroms.}
\tablenotetext{c}{The observed total line/multiplet flux in units of
$10^{-6}$ photons cm$^{-2}$ s$^{-1}$ and the associated $1\sigma$
errors.}
\tablenotetext{d}{The total number of counts in all components of a
spectral line (continuum subtracted)}
\tablenotetext{e}{The total flux from the strongest component
$\lambda\lambda 15.01, 15.26$~\AA~and~
$\lambda\lambda\lambda 16.78, 17.05, 17.10$~\AA, respectively.}
\end{center}
\end{table}

%% If the table is more than one page long, the width of the table can vary %
%from page to page when the default \tablewidth is used, as below.  The %
%individual table widths for each page will be written to the log file; a %
%maximum tablewidth for the table can be computed from these values.  % The
%\tablewidth argument can then be reset and the file reprocessed, so % that
%the table is of uniform width throughout. Try getting the widths % from the
%log file and changing the \tablewidth parameter to see how % adjusting this
%value affects table formatting.

%% The \dataset macro has also been applied to a few of the objects to % show
%how many observations can be tagged in a table.

\end{document}